\input{aipcheck}

\documentclass[
    ,final            
  ]
  {aipproc}

\layoutstyle{6x9}

\newcommand{\hi}{H{\sc i}}
\newcommand{\ha}{H{$\alpha$}}
\newcommand{\kms}{km\,s$^{-1}$}

\begin{document}

\title{Connecting Gas Dynamics and Star Formation Histories in Nearby
  Galaxies: The VLA--ANGST Survey}


\author{J\"urgen Ott\footnote{J\"urgen Ott is a Jansky Fellow of the National Radio Astronomy Observatory; email: jott@nrao.edu}}{
  address={National Radio Astronomy Observatory, 520 Edgemont Road, Charlottesville, VA 22903, USA}, altaddress={California Institute of Technology, 1200 E California Blvd, Caltech Astronomy 105-24, Pasadena, CA 91125, USA}
}

\author{Evan Skillman}{
  address={Department of Astronomy, University of Minnesota, 115 Church St. SE, Minneapolis, MN 55455, USA}
}

\author{Julianne Dalcanton}{
  address={Department of Astronomy, University of Washington, Box 351580, Seattle WA 98195, USA}
}

\author{Fabian Walter}{
  address={Max-Planck-Institute for Astronomy, K\"onigsstuhl 17, 69117 Heidelberg, Germany}
}

\author{Adrienne Stilp}{
  address={Department of Astronomy, University of Washington, Box 351580, Seattle WA 98195, USA}
}

\author{B\"arbel Koribalski}{
  address={Australia Telescope National Facility, CSIRO, PO Box 76, Epping, NSW 1710, Australia}
}

\author{Andrew West}{
  address={Department of Astronomy, University of California, 601 Campbell Hall, Berkeley, CA 94720, USA}
}

\author{Steven Warren}{
  address={Department of Astronomy, University of Minnesota, 115 Church St. SE, Minneapolis, MN 55455, USA}
}

\begin{abstract}
  In recent years, HST revolutionized the field of star formation in
  nearby galaxies. Due to its high angular resolution it has now
  become possible to construct star formation histories of individual
  stellar populations on scales of a few arcseconds spanning a range
  of up to $\sim 600$\,Myr. This method will be applied to the ANGST
  galaxies, a large {\it HST} volume limited survey to map galaxies up
  to distances of 3.5-4.0\,Mpc (excluding the Local Group). The ANGST
  sample is currently followed--up by high, $\sim 6''$ resolution VLA
  observations of neutral, atomic hydrogen (\hi) in the context of
  VLA-ANGST, an approved Large VLA Project. The VLA resolution is well
  matched to that of the spatially resolved star formation history
  maps. The combination of ANGST and VLA-ANGST data will provide a
  new, promising approach to study essential fields of galaxy
  evolution such as the triggering of star formation, the feedback of
  massive stars into the interstellar medium, and the structure and
  dynamics of the interstellar medium.

%
%
\end{abstract}

\maketitle


\section{Introduction}

Luminous matter in galaxies is being converted from the interstellar
matter (ISM) phase into stars via the processes of star formation
(SF). In turn, young, massive stars are short lived and release energy
and matter back into the ISM in the form of strong stellar winds and
supernova explosions, the so--called feedback processes. The detailed
interactions of gas and stars are complex and it is not straight
forward to predict the evolution of gas rich galaxies from any given
state. The ISM must meet certain conditions to allow gravitational
collapse of the material and eventually to form a star. But what
exactly these conditions are and how they develop remains
unclear. Many different models have been proposed to trigger star
formation, e.g., stochastic modes \cite{sei79,kau06}, where stars form
out of gas density enhancements that are created by the random
velocity structure of the gas, external triggers such a tidal
interaction with companion galaxies or ram pressure effects due to the
galaxy's motion through an intergalactic medium \cite{kee93,fuj98},
turbulent SF \cite{lar81,bur02} where the turbulent ISM creates gas
density enhancements suitable for gravitational collapse, episodic SF
or 'breathing' modes \cite{stin07,qui08} where, after an initial
starburst strong feedback from massive stars inhibit subsequent SF
until the energy has dispersed and the gravitational pull of the
galaxy accumulates material to start another episode of SF
\cite{stin07,qui08}, or propagating SF where energetic feedback from
massive stars compresses the surrounding gas, leading to subsequent SF
events propagating away from the origin \cite{ost81,mcc87}. It remains
unclear what the different SF mechanisms contribute to the overall SF
history (SFH) of galaxies and for a full understanding is
indispensable to observe both, the gaseous and stellar content of
galaxies. This is the goal of our project. ANGST (ACS Nearby Galaxy
Survey Treasury, see next section) is an ambitious program to
determine the stellar content and spatially resolved SFHs in
galaxies. VLA-ANGST is a Very Large Array (VLA) follow-up on the ANGST
sample to map the neutral, atomic \hi\ gas content at high spatial and
high velocity resolution. Together, these surveys will provide
well-matched, rich datasets which will be used to examine SF theories.

\section{ANGST: ACS Nearby Galaxy Survey Treasury}
\label{sec:angst}

ANGST (PI: J.  Dalcanton, \cite{dal07}) is a large project to obtain
optical imaging in two bands (F814W, F475W) with the high resolution
capabilities of the Advanced Camera for Surveys (ACS) on board
of {\it HST}. It is a volume limited survey of all galaxies out to
distances of $\sim 3.5$\,Mpc, at Galactic latitudes
$|b|>20^{\circ}$. Due to their very large extent on the sky and the
relatively small field of view of {\it HST}, Local Group galaxies are
not part of the survey. To include most of the galaxies in the nearby
M\,81 and Sculptor groups, however, cones out to 4\,Mpc toward those
regions have been added to the survey volume. ANGST contains 69
galaxies of all morphological types and spans a wide range of $\sim
10$ magnitudes in optical luminosity and $\sim 4$ magnitudes of SF rate.

One method to derive the SFH of a galaxy is based on the luminosity
function of its main sequence. Every bin along the main sequence
contains stellar populations of a given age but, unfortunately, it
also contains younger stars which makes the interpretation less
reliable and limits the look-back time to $\sim 300$\,Myr. A better
method to disentangle the different stellar ages is to use the blue
loop of the stellar isochrones, occupied by helium burning stars -- a
method pioneered by \cite{doh97,doh98}. Not only can a \emph{global}
SFH up to $\sim 600$\,Myr be derived with this method but the quality
of the {\it HST} observations allows to pin down stars of a given age
(time resolution up to $\sim 30$\,Myr) and it is thus possible to
create {\it maps} of the resolved SFH of galaxies \cite{doh02}.

\section{VLA--ANGST}

Using the Very Large Array in B, C, and D configuration, VLA-ANGST
maps the \hi\ content at high linear ($\sim 6''$) and velocity
resolution (a few \kms). This is well matched to the spatial
resolution that can be achieved in the construction of the resolved
SFH for the ANGST targets. Not all of the ANGST galaxies are observed
in VLA-ANGST. Objects too far south for the VLA as well as galaxies
with known \hi\ non-detections are excluded. A few of the ANGST targets
have already been mapped by the 'The \hi\ Nearby Galaxy Survey'
(THINGS, PI: F. Walter \cite{wal08}), an \hi\ survey with the exact same
observational setup as VLA-ANGST, and the data is shared. In total,
VLA-ANGST has an allocated time of $\sim 480$\,hours of VLA time and
the data collection is expected to be finished by the end of 2008.

\begin{figure}
  \includegraphics[width=\textwidth]{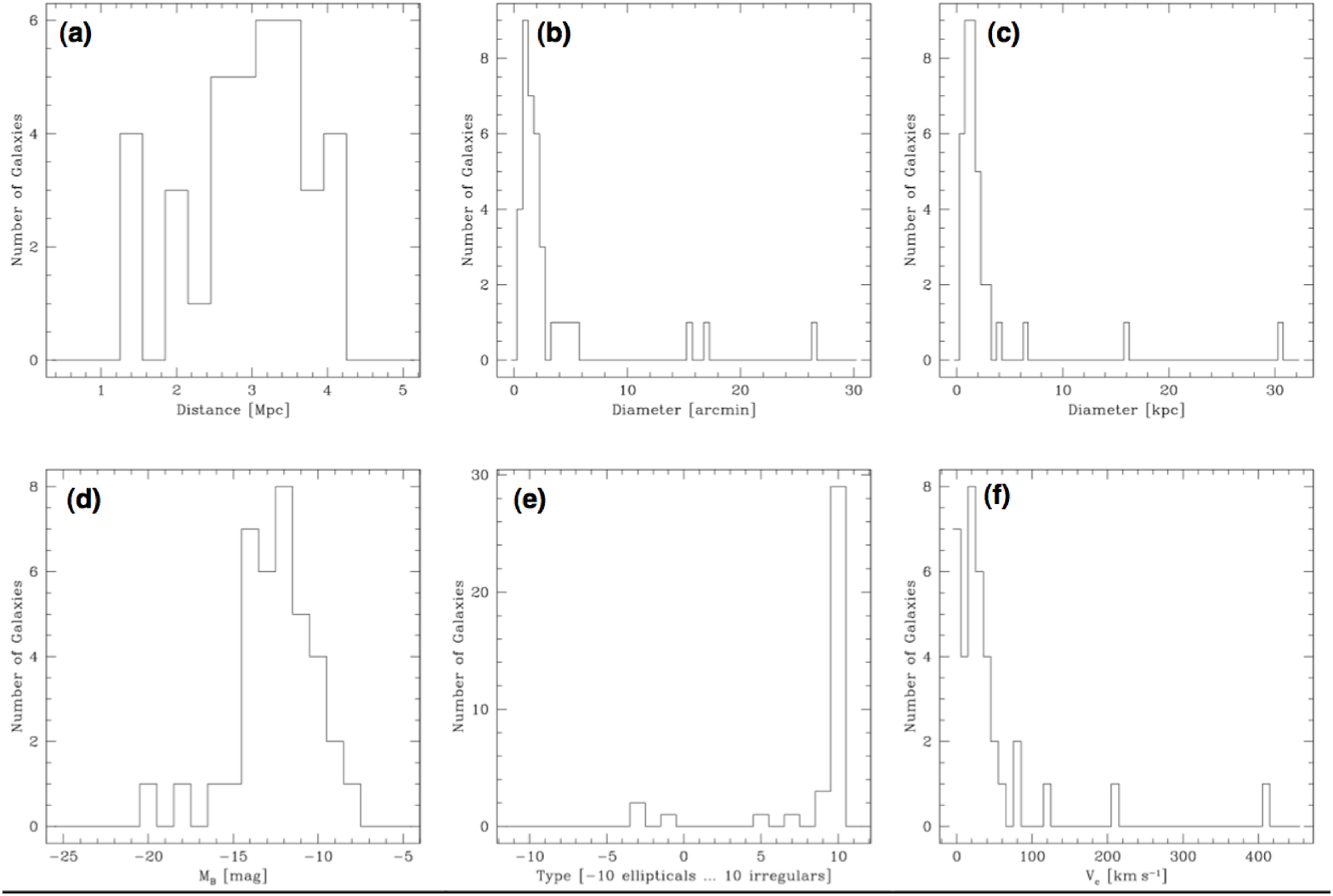}
  \caption{Properties of the VLA-ANGST sample: Histograms of {\bf (a)} distance, {\bf (b)} apparent diameter, {\bf (c)} physical diameter, {\bf (d)} absolute blue magnitude, {\bf (e)} galaxy type (see Table\ref{tab:sample} for nomenclature), and {\bf (f)} circular velocity.\label{fig:sample}} 
\end{figure}

The VLA-ANGST sample is listed in Table\,\ref{tab:sample} and
histograms of the number of galaxies as functions of galaxy parameters
are shown in Fig.\,\ref{fig:sample}. Most of the galaxies are at a
distance of $\sim 3$\,Mpc, where the M\,81 and the Sculptor groups
reside. The size distribution peaks at small diameters; most galaxies
exhibit rather faint absolute blue magnitudes of $\sim
-13$\,mag. Together with their morphological classification
(Fig.\,\ref{fig:sample}[e]) it becomes obvious that the sample is
dominated by gas rich dwarf irregular galaxies (dIrrs). This is
expected for a volume limited survey, because dwarfs are the most
numerous type of galaxy, and because dwarf spheroidals and ellipticals
are mostly excluded from VLA-ANGST due to the criterion to discard
ANGST galaxies with \hi\ non--detections. The dominance of dIrrs is
not unwelcome. As shown in Fig.\,\ref{fig:sample}(f), most of the
galaxies in our sample are rather slow rotators and are expected to be
largely governed by solid body motion. Shearing forces that may mix
stellar populations and smear out \hi\ structures are therefore much
less dominant and are usually found at the outskirts of dIrrs. The
comparison of stars and gas features are thus much more accurate and
can be traced back to older populations.

\begin{table}
\begin{tabular}{lrrrrrrr}
\hline
Name & D &    RA (J2000)  &       DEC(J2000) &    d  &    M$_{\rm B}$ &  Type  &   V$_{\rm c}$\\ 
&Mpc &$h:m:s$ & $^{\circ}$:':'' &arcmin &mag& &km\,s$^{-1}$\\ 
\hline
\multicolumn{8}{c}{Galaxies with {\sc Hi} detections and recent SF, brighter than -13}\\


N3109	&1.3	&10:03:07.2	&-26:09:36	&17.0  &-15.18	&9	&116\\ 
SexA	&1.3	&10:11:00.8	&-04:41:34	&5.5   &-13.71	&10	&63\\ 
SexB	&1.4	&10:00:00.1	&05:19:56	&5.1   &-13.88	&10	&38\\ 
DDO125	&2.5	&12:27:41.8	&43:29:38	&4.3   &-14.04	&10	&30\\ 
DDO99	&2.6	&11:50:53.0	&38:52:50	&4.1   &-13.37	&10	&37\\ 
DDO190	&2.8	&14:24:43.5	&44:31:33	&1.8   &-14.14	&10	&45\\ 
N3741	&3.0	&11:36:06.4	&45:17:07	&2.0   &-13.01	&10	&81\\ 
N4163   &3.0    &12:12:08.9     &36:10:10       &1.9   &-13.76  &10     &18\\ 
N4190	&3.5	&12:13:44.6	&36:38:00	&1.7   &-14.20	&10	&46\\ 
N247	&3.6	&00:47:08.3	&-20:45:36	&15.4  &-17.92	&7	&210\\ 
N253	&3.9	&00:47:34.3	&-25:17:32	&26.7  &-20.04	&5	&410\\ 
\hline

\multicolumn{8}{c}{Galaxies with HI detections and recent SF, fainter than -13\,mag}\\

Antlia &1.3 &10:04:04.0 &-27:19:55  &2.0 & -9.38  &10 &22\\ 
KK230  &1.9 &14:07:10.7 &35:03:37   &0.6 & -8.49  &10 &21\\ 
GR8    &2.1 &12:58:40.4 &14:13:03   &1.1 & -12.00 &10 &26\\ 
DDO187 &2.3 &14:15:56.5 &23:03:19   &1.7 & -12.43 &10 &34\\ 
KKH98  &2.5 &23:45:34.0 &38:43:04   &1.1 & -10.29 &10 &22\\ 
U8508  &2.6 &13:30:44.4 &54:54:36   &1.7 & -12.95 &10 &49\\ 
DDO181 &3.0 &13:39:53.8 &40:44:21   &2.3 & -12.94 &10 &39\\ 
UA292  &3.1 &12:38:40.0 &32:46:00   &1.0 & -11.36 &10 &27\\ 
U8833  &3.1 &13:54:48.7 &35:50:15   &0.9 & -12.31 &10 &28\\ 
U4483  &3.2 &08:37:03.0 &69:46:31   &1.2 & -12.58 &10 &33\\ 
DDO6   &3.3 &00:49:49.3 &-21:00:58  &1.7 & -12.40 &10 &22\\ 
KKH37  &3.4 &06:47:45.8 &80:07:26   &1.2 & -11.26 &10 &20\\ 
KDG73  &3.7 &10:52:55.3 &69:32:45   &0.6 & -10.75 &10 &18\\ 
HS117  &4.0 &10:21:25.2 &71:06:58   &1.5 & -11.51 &10 &13\\ 
BK3N   &4.0 &09:53:48.5 &68:58:09   &0.5 & -9.23  &10 &15\\ 
\hline

\multicolumn{8}{c}{Galaxies with HI detections but little evidence for recent SF}\\


KKR25	&1.9	&16:13:47.6	&54:22:16	&1.1  &  -9.94&	10&    15\\ 
KKH86	&2.6	&13:54:33.6	&04:14:35	&0.7  & -10.19&	10	&14\\ 

\hline
\multicolumn{8}{c}{Galaxies with HI detections and dE morphology}\\

N404 & 3.1	&01:09:26.9&	35:43:03&	2.5&   -16.25&	-1&	78\\ 
KDG63&	3.5	&10:05:07.3&	66:33:18&	1.7&   -11.71&	-3&	19\\ && y & y & y \\
\hline
\multicolumn{8}{c}{Galaxies with no reported single-dish detection/observation, but dIrr/Sm morphology} \\

DDO113	     &2.9	&12:14:57.9	&36:13:08	&1.5   &-11.61	&10	&?\\ 
DDO183      &3.2        &13:50:51.1     & 38:01:16      &2.2   &-13.08	&9	&?\\ 
MCG9-20-131 &3.4	&12:15:46.7	&52:23:15	&1.2   &-12.36	&10	&?\\ 
A0952+69    &3.9	&09:57:29.0	&69:16:20	&1.8   &-11.16	&10	&?\\ 
DDO82	     &4.0	&10:30:35.0	&70:37:10	&3.4   &-14.44	&9	&?\\ 

\hline

\multicolumn{8}{c}{Galaxies with no reported single-dish detection/observation and dE morphology, but recent SF}\\

KK77&	3.5&	09:50:10.0&	67:30:24&	2.4&   -11.42&	-3&	?\\ 
\hline

\end{tabular}
\normalsize
\caption{VLA-ANGST targets. D: distance, d:
  diameter, M$_{\rm B}$: absolute blue magnitude, Type: galaxy type with
  $<0$ = elliptical, Sb=3, Sc=5, $>7$=Sd or later, V$_{\rm c}$: circular
  velocity defined as W$_{50,c}/2$). \label{tab:sample}}
\end{table}


\subsection{VLA-ANGST Science Goals}

The combination of ANGST and VLA-ANGST will provide new tools to
understand the interplay between stars and gas in galaxies. In
particular, the surveys will contribute to answer the following
questions:\\


\noindent {\it What causes gas to collapse into stars?  What are the triggers of star formation?}\\
The \hi\ data will be compared to maps of resolved SFH. Both the
distribution as well as the dynamics of the gas are expected to play
key roles in defining the actual sites where star formation can
occur. Different star formation models (see Introduction) lead to
different properties of the gas in their relation to stellar contents
and ages. SF efficiencies will be derived which may depend on the SFH
of a galaxy, on its morphology, and/or on its metallicity. The high
resolution of our data will be reflected in precision rotation curves,
which is indispensable in understanding how shearing influences SF,
and to identify regions where the galaxies spin like solid
bodies. Those areas are the most promising
ones to test stochastic and propagating SF modes in galaxies.\\

\noindent {\it What is the role, efficiency, and impact of stellar
  feedback on the ISM?}\\
With the stellar population age and mass distribution at any given
position in a galaxy, the kinematic properties of the \hi\ gas can be
used to estimate what fraction of the energy released by strong
stellar winds and supernova explosions is dumped into the ISM. This
will also help to better understand the mechanism of expanding \hi\
supershells, which are thought to be a signature of feedback from
entire stellar clusters \cite{ten88}, but for many of the shells the
search for such a cluster has failed \cite{rho99}. When regions with
and without strong shear are compared, it will be possible to
determine how much of the \hi\ turbulence is caused by external,
galaxy wide dynamics and what fraction originates from more localized
SF events. The energetic input from massive stars can have two very
different effects, it can either compress the surrounding gas, causing
it to collapse and to expedite further SF, or it can stir up the
surrounding ISM in a way such that SF is suppressed. Our data will
make it possible to derive which of the processes is more dominant.\\

\noindent {\it What is the structure and dynamics of the
  ISM?}\\
For many nearby galaxies, a wealth of data across all wavelengths will
become available in the very near future. Surveys, e.g., the 11HUGS
(``The 11 Mpc \ha\ UV Galaxy Survey'', PI: J. Lee \cite{lee07}), SINGS
(``{\it Spitzer} Nearby Galaxy Survey'', PI: R. Kennicutt
\cite{ken03}), LVL (``{\it Spitzer} Local Volume Legacy'', PI:
R. Kennicutt \cite{ken07}), or STING (``Survey Toward Infrared-bright
Nearby Galaxies'', CARMA CO maps, PI: A. Bolatto) add observations of
star formation tracers such as \ha, UV, FIR, and CO. The combination
of the multi--wavelength information will eventually lead to the
detailed understanding of the energy balance between the different
phases of the ISM, e.g., at which \hi\ densities and dispersions the
different SF tracers appear and how this bootstraps on SF
processes. In addition, the fractal properties of \hi, e.g., the
fractal dimension can be studied over
a large number of different galaxies with unified data quality.\\

\noindent {\it How do galaxies evolve?}\\
Gas is consumed by SF and feedback processes of massive stars are
replenishing part of the ISM. The knowledge of the global SFH of the
galaxies in our sample will allow us to derive the \hi\ to stellar
mass ratio as a function the cumulative SF rate over the galaxies'
lifetimes. Even better, the spatial resolution of the SF rate maps
enables us to derive variations of this ratio over the bodies of the
galaxies, e.g, in the cores and the outer disks. This will provide a
measure for the efficiency with which gas is turned into stars. In
addition, properties of the ISM will be brought in relation to the
type of their hosts. Large scale dynamics, e.g., bars and density
waves, will have an effect on the gas and SF properties. This
influences the evolution of a galaxy as a whole and the timing between
gaseous features (via kinematics) and stellar ages will be
derived. Rotation curves, which are needed for such a study and which
will be a product of VLA-ANGST, are also a gold mine to test $\Lambda$
cold dark matter ($\Lambda$CDM) models. The spatial resolution of the
data is good enough to derive the shape of the dark matter halo down
to the very cores of the galaxies. Whereas theoretical $\Lambda$CDM
models predict a cuspy profile, observations so far show a flat
density distribution in contradiction to $\Lambda$CDM
\cite{deb01}. The VLA-ANGST survey, in particular combined with the
THINGS and 'Little THINGS' surveys will provide a large, high spatial
and velocity resolution database of \hi\ in galaxies with very similar
data quality. $\Lambda$CDM tests on the shape of dark matter density
profiles will therefore have the excellent statistics of a 100+ nearby
galaxy sample.

\section{Outlook}

At this moment in time, a large number of interferometric \hi\ surveys
of nearby galaxies are underway or have just been finished. In the
south, the ``Local Volume \hi\ Survey''
(LVHIS\footnote{http://www.atnf.csiro.au/research/LVHIS}, PI:
B. Koribalski) is covering all galaxies with \hi\ out to $\sim
10$\,Mpc, with a spatial resolution of $\sim 20"$, the highest
economically feasible resolution of the Australia Telescope Compact
Array\footnote{The Australia Telescope Compact Array is part of the
  Australia Telescope which is funded by the Commonwealth of Australia
  for operation as a National Facility managed by CSIRO.}. In the
northern hemisphere, there is the Giant Meterwave Radio Telescope
``Faint Irregular Galaxies GMRT Survey'' (FIGGS, PI: A. Begum) as well
as three major high resolution ($\sim 6"$) VLA \hi\ surveys,
THINGS\footnote{http://www.mpia.de/THINGS} (``The \hi\ Nearby Galaxies
Survey'', PI: F. Walter),
VLA--ANGST\footnote{http://www.cv.nrao.edu/~jott/VLA-ANGST} (PI:
J. Ott), and 'Little
THINGS'\footnote{http://www.lowell.edu/users/dah/littlethings} (PI:
D. Hunter), ongoing or have recently been completed. All of the three
VLA surveys are designed to provide data very similar in spatial
resolution (VLA B, C, and D array configurations), velocity resolution
(a few \kms), and sensitivity (\hi\ column density limit $\sim
10^{19}$\,cm$^{-2}$). In addition, the data reduction of the three
surveys will be very similar. This will provide almost equally high
fidelity imaging for a total of more than a hundred
galaxies. Extragalactic high resolution \hi\ studies will therefore
enter a new era with statistically meaningful results. Together with
surveys at other wavelengths that are currently underway (ANGST, LVL,
11HUGS, STING, etc.), we will have all the tracers at hand to
understand the interplay of gas and stars on scales of galaxies and
galaxy groups. This will lead to hard tests for SF and galaxy
evolution theories and substantially further our understanding of the
processes involved.

%
\begin{theacknowledgments}
The National Radio Astronomy Observatory is a facility of the National Science Foundation operated under cooperative agreement by Associated Universities, Inc.
\end{theacknowledgments}

\end{document}